\documentclass[a4paper,reviewcopy]{elsart}
\usepackage{graphicx,amsmath,amssymb,amsfonts}

\journal{Computer Physics Communications}
\date{30 October 2007}

\newcommand{\python}{\textsc{Python}}

\begin{document}

\begin{frontmatter}


\title{Evolutionary Fitting Methods for the Extraction of
  Mass Spectra in Lattice Field Theory}

\author{Georg M. von Hippel\thanksref{desy}}
\thanks[desy]{new address: DESY, Platanenallee 6, 15738 Zeuthen, Germany.},
\author{Randy Lewis}, and
\author{Robert G. Petry}.

\address{Department of Physics, University of Regina, Regina,
  SK, S4S 0A2, Canada.}

% Abstract

\begin{abstract}

  We present an application of evolutionary algorithms to the
  curve-fitting problems commonly encountered when trying to extract
  particle masses from correlators in Lattice QCD. Harnessing the
  flexibility of evolutionary methods in global optimization allows us
  to dynamically adapt the number of states to be fitted along with
  their energies so as to minimize overall $\chi^2$/(d.o.f.), leading to
  a promising new way of extracting the mass spectrum from measured
  correlation functions.

\end{abstract}

% Keywords, PACS and MSC codes


\end{frontmatter}

% Main text

%
%
%
%
%
\section{Introduction}
\label{sec_intro}

Curve fitting plays a central role in the analysis of lattice
simulation data. One of the most important and common uses of
curve fitting in lattice gauge theory is the extraction of
hadronic masses and matrix elements from measured correlation
functions.

The problem of extracting the mass spectrum from the
correlators measured in Lattice QCD simulations is well known
\cite{Lepage:2001ym,Lin:2007iq}.
The simulation produces data $\overline{G_i}=\overline{G(t_i)}$ for the
expectation values of the correlator $G(t)$ at a finite number of
discrete (Euclidean) time steps $t_i$, $1\leq i\leq N$. On the other hand,
from theory the exact form of the propagator is known to be given by
an infinite series\footnote{For periodic boundary conditions,
  hyperbolic functions will appear instead of the exponential. For a
  static particle, the energy is simply the particle mass.}
\begin{equation}
\label{eqn:thform}
G(t) = \sum_{n=0}^\infty Z_n \e^{-E_n t}\;,
\end{equation}
where we will assume that the energy levels are ordered, $E_n\leq
E_{n+1}$. The problem is then to determine an infinite number of
amplitudes $Z_n>0$ and energies $E_n>0$ from only a finite number of data
points $\overline{G_i}$, an obviously ill-posed problem.

To make the problem well-posed, we have to add some further physical
constraints. The piece of theoretical information that is normally
used is that the sequence of the $Z_n$ is bounded from above, and
therefore the correlator will be dominated by the lowest few terms at
all but the smallest values of $t$ due to the exponential suppression
by $E_n$. We can therefore truncate the series in equation~(\ref{eqn:thform})
after a finite number of terms, provided we only
attempt to fit at $t>t_{min}$ large enough for the truncation to
include all terms that make a significant contribution.

In doing so, we are faced with a choice: Either take a fixed
number $n_{max}$ of terms of the sum in (\ref{eqn:thform}) and adjust
$t_{min}$  so as to obtain a good fit, or fix $t_{min}$ and vary
$n_{max}$ so as to extract the largest possible amount of significant
information. The problem with the first strategy is that we are
throwing away valuable information contained in the data points at
$t<t_{min}$, leading to large statistical errors if $t_{min}$
is chosen too big.
These need to be balanced against large systematic errors arising if
$t_{min}$ is chosen too small for the given $n_{max}$\nolinebreak
\cite{Lepage:2001ym}.
In this paper we will therefore adopt the second approach and attempt
to fit all data points (excluding only the single point at $t=0$ for
practical reasons) with a variable number $n_{max}$ of exponentials.

Naively, one might want to try to simply run a series of fits using a
state-of-the art fitting method such as Levenberg-Marquardt
\cite{Levenberg:1944,Marquardt:1963}
at a number of different $n_{max}$, and with a variety of initial parameter
values, and choose the fit that produces
the lowest $\chi^2$ per degree of freedom. This method can be
made to work in the case of one single correlator if the problem of
finding an appropriate starting point in a potentially multi-modal
landscape is somehow solved.  However, for many questions in lattice QCD it is
necessary to fit multiple correlators, which may or may not share some
of the energy levels $E_n$. In this case, the fast combinatorial
growth with $n_{max}$ of the number of possibilities of assigning
shared or separate fit parameters $E_n$ to the fitting functions for
the different correlators renders the application of this naive method
to those problems largely impossible.
The problem of choosing appropriate initial parameter values further
exacerbates this approach.

The current state of the art using the variable-$n_{max}$ approach is
that taken in~
\cite{Lepage:2001ym}
where it is used in the context of a Bayesian method
\cite{Morningstar:2001je}
of \emph{constrained curve fitting}. The latter works by adding
prior information about the fit parameters and using it to constrain
the fit to a smaller subset of likely parameter values out of the
space of all possible values. Only those parameters whose fitted
values are largely independent of the priors used to constrain them
are considered to be determined by the data, while the others are
disregarded as having been imposed by the priors.

While it is thus possible to determine which
fitted quantities are independent of, or only weakly dependent on, the
priors and thus determined from the data, the idea of using some
external knowledge as an input could be seen as
incompatible with the notion of a first-principles determination of
the quantities of interest from QCD itself, without any recourse
whatsoever to empirical models. Moreover, in some cases appropriate
priors may not be available. Under those circumstances, it becomes
desirable to be able to extract some estimate of the parameters
to be fitted from the data themselves, and to use this estimate as a prior
in the context of a Bayesian constrained fit. A number of methods to
do this, such as the \emph{sequential empirical Bayes method}
\cite{Chen:2004gp},
have been used in the existing literature.

A completely different state-of-the-art approach that does not rely on
prior information while allowing for extraction of a spectrum from
multiple sets of data for improved statistics is
the \emph{variational method}
\cite{Michael:1985ne,Luscher:1990ck,Burch:2005wd}.
Here one sets out by fixing a channel corresponding to a
specific set of quantum numbers, and finds a set of appropriate
linearly independent operators $O_i$ for the channel. One then
calculates all diagonal $\langle O_iO_i\rangle$ and off-diagonal
$\langle O_iO_j\rangle,~i\not=j,$ correlators between these operators, each
operator having a corresponding form for the sink that annihilates the
state created by the operator at the source.  One may then use the
variational method on this cross-correlator matrix $C_{ij}(t)$ to find a
superposition of the original operators that corresponds to the lowest
energy state of the channel.  Assuming a meson channel, the diagonal
correlator of this operator may be fit by a single exponential
function whose mass in the exponent is the ground state
energy. Operators corresponding to excited states, whose diagonal
correlators ideally correspond to functions only of the excited state
masses, may be produced in a similar manner
\cite{Morningstar:1999rf}.

While very powerful, the variational method has a number of features
that limit the scope of its applicability. Firstly, the use of
the variational method is facilitated by the selection of operators
such that the correlator matrix is hermitian
\cite{Luscher:1990ck,Basak:2005aq}.
However, in lattice simulations it is often desirable to suppress
unwanted but usually present contributions from high-frequency modes
by means of \emph{quark smearing} of the operators
\cite{Alford:1995dm,Lepage:1989hd,Basak:2005gi}.
In this case, hermiticity of the correlator matrix requires smearing
to be applied both at the source and the sink of the correlator, which
in most cases is very expensive computationally.
Secondly, the variational approach also requires both source and sink
operators to have the exact quantum numbers of the channel of
interest, instead of the less stringent usual requirement that both
operators overlap with the state of interest and that at least one of
them have its exact quantum numbers. This requirement limits
the selection of correlators that are usable with the variational method.
Finally, the variational method requires all off-diagonal correlators
to be calculated. The number of needed correlators thus grows as $N^2$,
with $N$ the number of operators in the channel of interest.
Since automated generation of group theoretical operators for a given
symmetry channel
\cite{Basak:2005aq,Lacock:1996vy,Basak:2005ir}
means that the number of operators that could be used may be rather
large, this quadratic growth in computational complexity imposes
further limitations. Thus correlator selection may be done more
judiciously without the restrictions imposed by the variational method.

A different problem arises in the case where one desires to fit correlators
corresponding to channels with \emph{different} lattice quantum
numbers.  An example of this occurs when the continuous
rotational symmetries of a quantum theory get broken when discretizing
it to a lattice. In the continuum, the rotational group has
representations corresponding to arbitrary integer or half-integer
angular momenta $J$, but on the lattice
one only retains the finite symmetry of the octahedral group.
A particle of angular momentum $J$ will therefore appear in one or
more of the lattice channels labeled by those irreducible
representations of the octahedral group to which the value of $J$
subduces
\cite{Johnson:1982yq}.
Thus a particle of integral spin will appear in some known subset of
the irreps $A_1$, $A_2$, $E$, $T_1$ and $T_2$ of the octahedral group.
Therefore, the
extraction of the physical spectrum is complicated by the fact that a
physical state's mass may lie in several lattice symmetry channels.
One needs to
identify in which irreps the physical state lies not only to aid in
identifying its physical angular momentum $J$ via its subduction
signature, but also to extract its mass as the latter is contained in the
data in all channels in which it appears.  Since a state either lies
within a lattice symmetry channel or it does not, the process of
state identification and fitting requires an algorithm that is inherently
discontinuous. As the number of resolvable states in lattice
simulations increases, the need for a systematic solution to this
second problem will as well.

Evolutionary fitting algorithms, while widely used in other areas of
research
\cite{Allanach:2004my,Ireland:2003yg,Mokiem:2005qf,Georgiev:2005af,%
Gozdziewski:2003sd,Gozdziewski:2003rv,Roncaratti:2005,Karr:1995},
are not currently in common use in lattice QCD.
In this paper, we present an application of evolutionary algorithms
\cite{Goldberg:1989,Reeves:2002,Whitley:1994,Whitley:2001}
to the problem of extracting mass spectra from hadronic correlators.
We believe that the advantages of evolutionary algorithms are
particularly pertinent to this problem: evolutionary algorithms allow
dynamic variation of the functional form of the fit function (such as the
number of states to be fitted) in a natural way so as to minimize
$\chi^2$ per degree of freedom;  the data themselves thus tell
us how many states can be reliably extracted. In addition,
evolutionary algorithms are inherently global optimizers and as such
largely eliminate any residual dependence of the result on the initial
guesses used as starting values, which may be a problem when using
conventional local optimizers.
Furthermore, based as they are on the Darwinian
principle of adaptation by mutation and natural selection
\cite{Darwin:1859},
evolutionary fitting methods excel at extracting information from data
without the need for any external prior information.
Finally, evolutionary fitting is able to fit multiple correlator datasets
with either identical or differing quantum numbers.

In this paper we aim to introduce lattice theorists to the
possibilities and features of evolutionary fitting algorithms and
to present some preliminary results from our practical
implementation of such an algorithm.
Section \ref{sec_evolalg} gives a general overview of the concepts and
ideas of evolutionary algorithms. Section \ref{sec_evsingle} details
an evolutionary algorithm that can be used to fit a single hadronic
correlator, and section \ref{sec_evmulti} outlines the adaptations to
the algorithm that are necessary in order to fit across multiple
datasets. Some possible variations of the basic algorithm are
explained in section \ref{sec_altvar}.

\section{Evolutionary Algorithms}
\label{sec_evolalg}

Evolutionary computing and genetic algorithms
\cite{Goldberg:1989,Reeves:2002}
are a very active field in computer science and numerical optimization
(cf. e.g. the review
\cite{Whitley:1994}
and references therein). By borrowing concepts from evolutionary
biology, evolutionary computing is able to harness the power of natural
selection by mutation and selective breeding for the purpose of
solving function optimization and related problems. The nomenclature
is borrowed from evolutionary biology as well: candidate solutions are
called ``organisms'', whose internal encodings are their
``genotypes'', the target function is referred to as ``fitness'', and each
step of the algorithm produces a new ``generation''.

A number of fine distinctions between ``genetic algorithms'',
``evolution strategies'', ``evolutionary programming'' and related
evolutionary algorithms and methods is sometimes employed in the
literature
\cite{Whitley:2001};
here we will use the term ``evolutionary algorithm'' broadly to mean
any global optimization method that relies on some form of random
mutation combined with selective breeding in a population of candidate
solutions
\cite{Whitley:1994}.
{}From this point of view, an evolutionary algorithm consists of the
following ingredients:
\begin{itemize}
\item A search space $\mathcal{G}$,
\item a fitness function $f:\mathcal{G}\rightarrow \Rset$, assumed to
  be bounded from above,
\item a mutation function $M_{\eta}:\mathcal{G}\rightarrow\mathcal{G}$, and
\item a selection function $S_{\eta'}:\mathcal{G}^N\rightarrow\mathcal{G}^N$
\end{itemize}
Here $N$ is the (fixed) size of the population.
Both the mutation and the selection function depend on some
uncorrelated white noise $\eta,\eta'$ that acts as a source of
randomness. The update step on a population $P_\tau\in\mathcal{G}^N$ is then
\begin{equation}
P_{\tau+1} = S_{\eta'}(M_\eta(P_\tau))\;.
\end{equation}
The mutation function acts on each individual in a population
separately, while the selection
function performs comparisons in fitness between individuals, and may
also involve crossover or ``sexual'' reproduction, which creates a new
individual from a pair of ``parent'' individuals.

The simplest selection procedure to use is straightforward ``elitist''
selection of the fittest $N_{elite}<N$ individuals from a population
of size $N$, followed by repopulation with their pairwise offspring
(where a child is formed by random interpolation between the parameter
values from either parent for each parameter). More sophisticated
selection procedures (such as ``roulette wheel'', ``rank'' or ``tournament''
selection
\cite{Whitley:1994})
could be used instead. Adding additional, less fit,
members to the elite on the basis of genetic distance may be helpful
in maintaining genetic diversity and avoiding premature convergence. 

If we require that the mutation function leaves the fittest individual
in a population alone, thereby ensuring
\begin{equation}
\label{eqn:nomutatefittest}
\max_{p\in M_\eta(P)} f(p) \ge \max_{p\in P} f(p)\;,
\end{equation}
and that the selection procedure never decreases the maximum fitness
within a population,
\begin{equation}
\label{eqn:fittestleftalone}
\max_{p\in S_\eta(P)} f(p) \ge \max_{p\in P} f(p)\;,
\end{equation}
it follows that the sequence $f_\tau=\max_{p\in P_\tau} f(p)$ is bounded
from above and monotonically increasing, and hence will converge.

Evolutionary algorithms are inherently global optimizers, as opposed to
local optimizers such as steepest descent or Newton methods. This
global nature is largely due to their inherent parallelism, since all
organisms in a population search the fitness landscape in parallel,
and also because crossover allows information learned by different
organisms to be combined and propagate throughout the population.
Another important feature of evolutionary algorithms is their ability
to optimize by varying discrete parameters (such as the functional
form of a candidate solution) in a natural and straightforward way,
something which local optimizers relying on certain smoothness
assumptions about the target function cannot easily do.

While evolutionary algorithms have not yet become a standard tool in
lattice QCD, they have previously been used for the purpose of
nonperturbative Landau gauge fixing
\cite{Markham:1998sf,Oliveira:2001kk,Oliveira:2003wa,Yamaguchi:1999hq}
and (in combination with an accept-reject step to achieve detailed
balance) as a simulation algorithm
\cite{Yamaguchi:1998yz,Yamaguchi:1999hp}.
In other fields of physics, evolutionary fitting algorithms have been
used among other things to discriminate between different SUSY models
\cite{Allanach:2004my},
to analyze resonances in $p(\gamma,K^+)\Lambda$ reactions
\cite{Ireland:2003yg},
to fit stellar spectra
\cite{Mokiem:2005qf},
to determine the mass loss rate of stellar winds
\cite{Georgiev:2005af},
to search for extrasolar planets
\cite{Gozdziewski:2003sd,Gozdziewski:2003rv},
to build diatomic potentials using a variational method
\cite{Roncaratti:2005},
and to solve black-box system characterization problems in engineering
\cite{Karr:1995}.
In most of these cases, just as in this work, it has been observed
that combining the evolutionary algorithm with a conventional (local)
optimization step gave the best results.

We believe that the flexibility and global nature of evolutionary
algorithms makes them an excellent tool for the purpose of curve
fitting, especially when the exact form of the fitting function (such
as the number of exponentials to use in our case) is subject to some
kind of data-dependent uncertainty.

\section{Evolutionary Fitting of a Single Correlator}
\label{sec_evsingle}

In this and the following section we present some details of our
version\footnote{For a practical implementation, we have chosen
\python\
\cite{python},
augmented by SciPy
\cite{scipy},
because of its object-orientation and the
flexible list and tuple types it natively provides.}
of an evolutionary algorithm for fitting correlation functions in
lattice QCD.

For the purpose of fitting a diagonal meson correlator\footnote{For
  baryonic and off-diagonal correlators, the functional form needs
  some adjustments. Nevertheless, the general method works the same in
  those cases.}, the search space is
\begin{eqnarray*}
\mathcal{G} = \Big\{G\in\mathcal{C}(\Rset) & \Big| &
  \exists\, n_{max}>0,Z_n>0,E_n>E_{n-1}>0:\; \\ 
  && G(t)=\sum_{n=0}^{n_{max}} Z_n \left(\e^{-E_n t}+\e^{-E_n (T-t)} \right)
\Big\},
\end{eqnarray*}
where $T$ is the temporal extent of the periodic lattice.
The fitness function is $f(G)=-\chi^2(G)/n_{dof}(G)$, where
the correlated $\chi^2$
\cite{Michael:1993yj,Michael:1994sz}
is
\begin{equation}
\label{eq:chi2_single}
\chi^2(G) = \sum_{t_i,t_j}( \overline{G_i} - G(t_i))
                   (\sigma^{-1})_{ij} (\overline{G_j} - G(t_j))\;,
\end{equation}
with the covariance matrix defined by
\begin{equation}
\sigma_{ij} = \overline{G_iG_j}-\overline{G_i}\;\overline{G_j}\;,
\end{equation}
and where
\begin{equation}
\label{eq:ndof_single}
n_{dof}(G) = (t_{max}-t_{min}+1) - 2 n_{max}
\end{equation}
is the number of degrees of freedom of the fit given by
$G$.\footnote{Where a fitness function $f:\mathcal{G}\rightarrow
  [0;1]$ is desired, functions such as $f(G) =
  \e^{-\chi^2(G)/n_{dof}(G)}$ or $f(G) = 1/(1+\chi^2(G)/n_{dof}(G))$
  can be used instead.}

An organism $G\in\mathcal{G}$ can therefore be represented by
a list of $n_{max}$ pairs~$(Z_n,E_n)$, and it is this
representation on which the mutation and breeding algorithms outlined
below work.

There are two kinds of mutation steps needed to search the entire
search space~$\mathcal{G}$. The first amounts to increasing or
decreasing $n_{max}$ of an individual organism.
When making these length mutations it is valuable to keep the sum
$\sum_n Z_n$ of all amplitudes fixed.\footnote{This may be
  accomplished, for instance, by sharing the amplitude of a removed
  term between the remaining amplitudes, and by decreasing the
  amplitudes of the existing exponentials so as to keep the overall
  amplitude fixed when adding a new term.}  One reason for this is
that the coefficient sum $G(0)=\sum_n Z_n$ can become relatively
stable across the population early on, so that changing $n_{max}$ by
simply dropping the pair $(Z_{n_{max}},E_{n_{max}})$ or adding a
random pair will tend to produce an unfit organism, regardless of
whether some potentially desirable organism of the new length exists.
Moreover, a situation can occur where spurious near-degenerate states
are kept because the change in overall amplitude from just removing a
single one of them increases $\chi^2$ more than is offset by the
simultaneous increase in $n_{dof}$ caused by the mutation.
The second type of mutation performs a random modification of a pair
$(Z_n,E_n)$. A natural way to implement this
is the addition of a pair of independent Gaussian random variables to
the original pair.

Since mutations can become somewhat disruptive of already accumulated
genetic information in the later stages of evolution, it may be useful
to make the rate of mutation dependent on $\chi^2/n_{dof}$ in such a
way as to increase the search area early on, before contracting it to
a more local search around the optima already found as the algorithm
converges.

Mutation steps can also be combined with a finite number of steps
of a local optimization routine (such as Levenberg-Marquardt).%
\footnote{Specifically, we take the functional form implied by the genotype
and do a  fixed number of steps of the Levenberg-Marquardt routine using
the genotype's parameter values as the initial values.  The new
genotype's values are set to the result.  Some random subset of the parameter
values can be kept fixed in the routine if desired.}
We have found that the addition of this latter step can greatly
accelerate convergence by giving beneficial mutations an improved
chance of survival.\footnote{This kind of strategy has sometimes been
  referred to as a \emph{hybrid genetic} or \emph{memetic algorithm}
  \cite{Moscato:1989}.}

Putting these ingredients together, we arrive at the following
mutation algorithm:
\begin{enumerate}
\item Generate a random number $p\in[0;1]$.
\item If $p<p^{len}_0 (1-\e^{-\alpha^{len}\chi^2/n_{dof}})$ (where $p^{len}_0$,
  $\alpha^{len}$ are tunable parameters):
  \begin{enumerate}
    \item Generate a random number $p'\in[0;1]$.
    \item If $p'<(1-1/n_{max})$, decrease $n_{max}$ by one, removing
      one randomly chosen exponential $(Z_n,E_n)$ from the fit, and
      redistributing the amplitude in $Z_n$ between the neighboring
      exponentials.
    \item Else increase $n_{max}$ by one and add a new, randomly
      generated $(Z_n,E_n)$ pair to the fit, decreasing the
      amplitudes of the pre-existing exponentials so as to keep the
      total amplitude fixed.
  \end{enumerate}
\item Generate a random number $p''\in[0;1]$.
\item If $p''<p_0^{parm}$ (where $p_0^{parm}$ is a tunable parameter):
  \begin{enumerate}
    \item Generate $n_{max}$ pairs of Gaussian deviates
      $(\Delta Z_n,\Delta E_n)$ with zero mean and standard deviation
      $\sigma=\sigma^{parm}_0 (1-\e^{-\alpha^{parm}\chi^2/n_{dof}})$ (where
      $\sigma^{parm}_0$ and $\alpha^{parm}$ are tunable parameters).
    \item Replace $(Z_n,E_n)$ by $(Z_n+\Delta Z_n,E_n+\Delta E_n)$,
      unless this would lead to a negative new value for $E_n$ or
      $Z_n$.
  \end{enumerate}
  \item Optionally perform a local (e.g. Levenberg-Marquardt) optimization
    on the fit with probability $p_0^{local}$.
\end{enumerate}

This mutation procedure depends on a number of tunable parameters. In
a number of numerical tests on a set of synthetic data, we found that
the point to which the algorithm ultimately converged did not depend
on the values of these parameters, indicating good stability of the
answer. The rate of convergence, on the other hand, did depend on the
particular parameters chosen, although the dependence was small for
``sensible'' parameter choices. Generally, $p_0^{len}$ should not be
too small, in order to explore the full solution space; we found that
$p_0^{len}=0.5$ worked well. $\alpha^{len}$ did not appear to have a
crucial influence on convergence, and was set to $\alpha^{len}=0.2$ in
subsequent runs. Since element mutations can be
rather disruptive of already achieved partial convergence,
$p_0^{parm}$ has to be chosen reasonably small; we found $p_0^{parm}=0.1$
a sensible choice. The convergence rate did not appear to strongly
depend on $\alpha^{parm}$ and $\sigma_0^{parm}$, which were set to
$\alpha^{parm}=0.1$ and $\sigma_0^{parm}=0.5$, respectively.
On the other hand, performing a local optimization step can never be
harmful, and $p_0^{local}$ should be set as large as computational
resources allow. It should be pointed out that the optimal choice of
parameters will likely depend on the particular fitting problem
investigated in each case, since the fitness landscapes can
conceivably look very different for different data.

The breeding or crossover function returns a child organism from two
parent organisms (par1 and par2), and works as follows:
\begin{enumerate}
\item Let $n_{max}^{child}=\max(n_{max}^{par1},n_{max}^{par2})$.
\item Generate  $n_{max}^{child}$ pairs of independent uniformly
  distributed random numbers $(x_i,y_i)\in[-\delta;1+\delta]$.
\item For $n<\min(n_{max}^{par1},n_{max}^{par2})$, choose the fit
  parameters of the child to be  $(Z_n^{child},E_n^{child}) = 
     (x_{n}Z_n^{par1}+(1-x_{n})Z_n^{par2},
      y_{n}E_n^{par1}+(1-y_{n})E_n^{par2})$; for other $n$,
      choose them equal to the longer parent's fit parameter values.
\end{enumerate}
The fit parameter values of the child organism are therefore chosen in
a hypercube spanned by the parent's fit values. Allowing for the
possibility of extrapolation instead of interpolation by introducing a
parameter $\delta$ is necessary to avoid rapid contraction towards
central points. In agreement with
\cite{Allanach:2004my},
we found $\delta=0.2$ sufficient to prevent this
contraction. ``Parthenogenesis'' or ``cloning'' of an existing
individual is possible by breeding an organism with itself.

Putting these elements together, we arrive at the following basic
evolutionary step to generate the next generation from a given
population:
\begin{enumerate}
\item All organisms except the fittest one are subjected to mutations
  according to the mutation algorithm stated above.
\item Selection and breeding are carried out as follows:
  \begin{enumerate}
  \item The fittest $N_{elite}$ organisms are selected.
  \item Another $N_{diversity}$ organisms are selected at random in
    order to maintain genetic diversity and avoid premature
    convergence.
  \item For each possible combination of the selected organisms
    (including those containing the same organism twice), a child
    organism is created according to the breeding algorithm above, and
    these child organisms form the next generation.
  \item $N_{mutant}$ copies of organisms from the elite are added to
    the population and are subjected to targeted mutations as a form
    of local search around the elite.
  \end{enumerate}
\end{enumerate}
By not subjecting the fittest organism to mutations and including
parthenogenesis in the breeding step, we ensure that inequalities
(\ref{eqn:nomutatefittest})
and
(\ref{eqn:fittestleftalone})
are fulfilled, and that thus the algorithm will eventually converge.

The basic evolutionary step is repeated until $\chi^2/n_{dof}$
converges below a chosen threshold $[\chi^2/n_{dof}]_{max}$,
or until a chosen maximum number $\tau_{max}$ of generations has been
exceeded with no improvement in the best overall genotype for the last
$\tau_{static}$ generations.

In order to be able to get a handle on the overall stability of the
evolutionary fit, and also as an aid in a possible parallelization, we
add a final wrinkle by partitioning the total population into
``islands'' of equal size, each of which forms a separate population
to which the basic evolutionary step is applied independently of the
other islands. To avoid individual islands with particularly
unfavorable starting conditions getting stuck, a weak coupling
between islands is introduced by replacing the least fit organism on
each island with a randomly chosen immigrant from a randomly chosen
island with probability $p^{mig}$ before carrying out the selection
step. We find that the lowest-lying states are identified rather
consistently across all islands fairly early on; only for the most
massive states discrepancies in number, energy and amplitude are found
between different islands, sometimes even in late generations,
indicating that the identification of these states is somewhat
uncertain.

Our main program thus employs a multi-island ecosystem for the
evolutionary optimization and then tries to improve on the results of
the best evolutionary fit by performing a Levenberg-Marquardt
optimization upon it. If sufficiently many islands are being used, it is
possible to do this as a constrained fit where desired, with the
average and standard deviation of the parameter values derived from
bootstrapping
\cite{efron82}
the best island fits being employed as priors.  Real
parameter errors may be determined by bootstrapping the data
configurations and rerunning the fit algorithm.  However, as this could
in principle produce different functional forms for different bootstrap
configurations, and as this procedure may be quite expensive, one may
choose instead to apply a Levenberg-Marquardt fit to the bootstrap
configurations using the final functional form obtained from the
evolutionary algorithm, as one would do in a conventional fit.

We have settled on the latter procedure for our implementation, and
have found that the final Levenberg-Marquardt fit always returned
within errors (as estimated by the bootstrapped Levenberg-Marquardt
procedure) of the evolutionary fit, indicating that the evolutionary
fit was already close to optimal. The errors from the bootstrapped
Levenberg-Marquardt fit were
usually (though not always, depending on how well genetic diversity
between islands was preserved in each case) comparable to those
estimated from the difference between the fits obtained on different
islands.\footnote{For the purpose of these single correlator tests,
we used a large ecosystem with $100$ islands.} In the following, we
use the error estimates from the final bootstrapped Levenberg-Marquardt
fit as our estimate of the error in the fitted parameters.

\begin{figure}
\includegraphics[height=\textwidth,angle=270,keepaspectratio=]{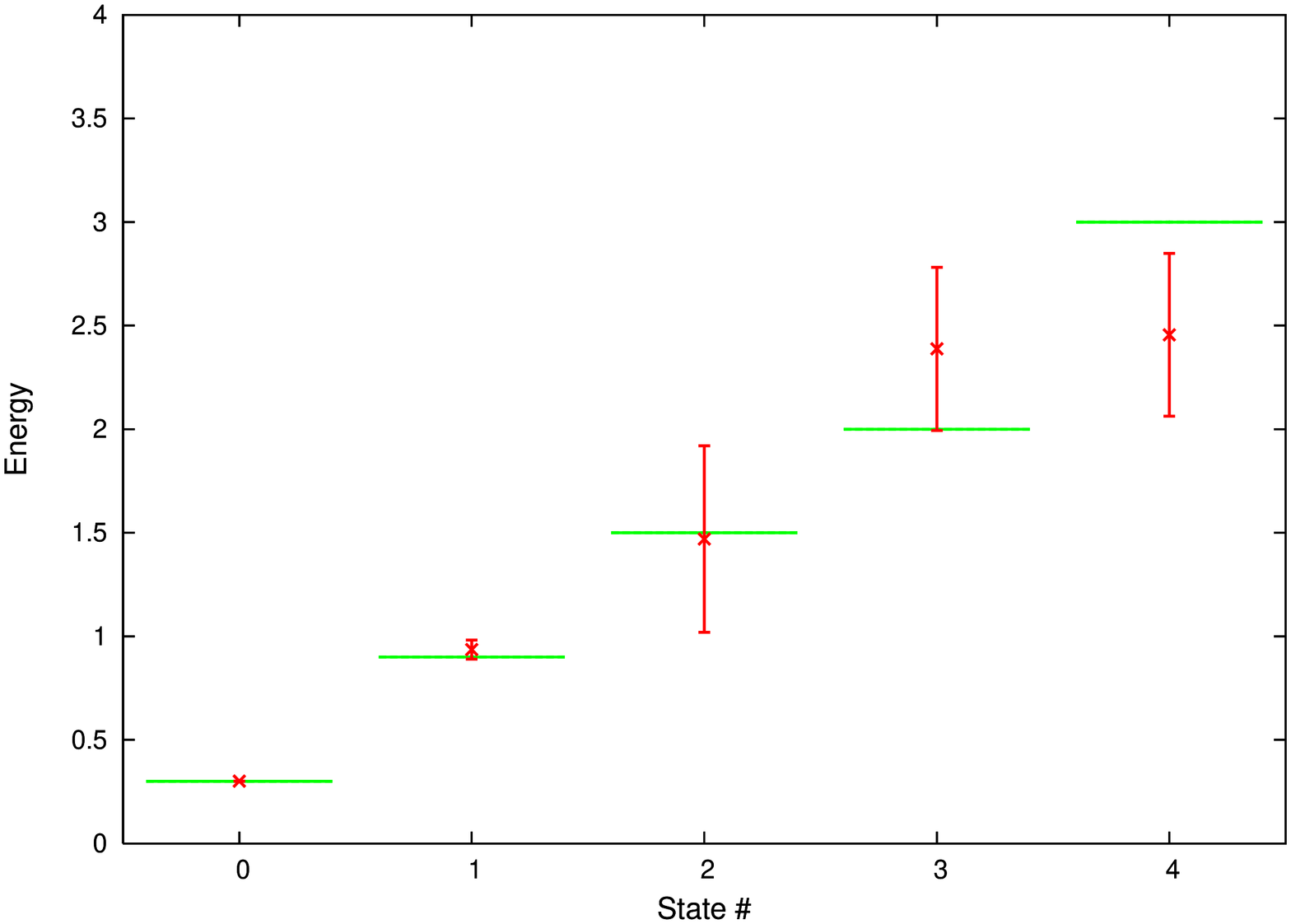}
\caption{Result of a fit to a set of pion-like synthetic
  correlators. Shown is a plot of the mass against the excitation
  number of the state. The horizontal lines are the input values
  used to create the synthetic data; the data points and errors
  are the fit results.\vspace{3ex}}
\label{fig:synthetic_pion}
\end{figure}
\begin{figure}
\includegraphics[height=\textwidth,angle=270,keepaspectratio=]{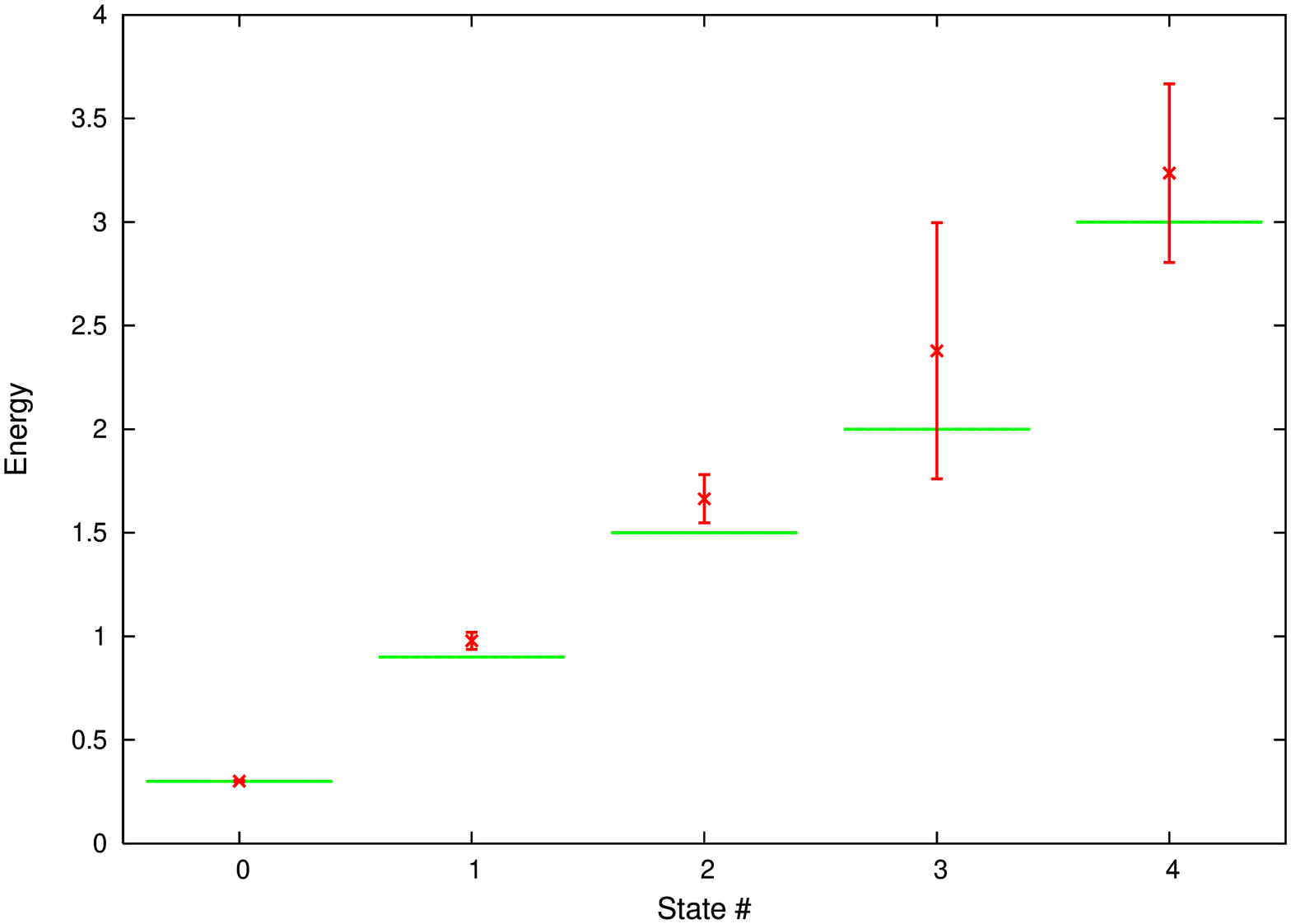}
\caption{Result of a fit to a set of rho-like synthetic
  correlators. Shown is a plot of the mass against the excitation
  number of the state. The horizontal lines are the input values
  used to create the synthetic data; the data points and errors
  are the fit results.\vspace{3ex}}
\label{fig:synthetic_rho}
\end{figure}

To demonstrate the viability of our method, we have run a fit on two
sets of synthetic data consisting of 200 artificial correlators for 48
timesteps, each constructed from a signal consisting of a sum of
exponentials with known masses and amplitudes, and Gaussian noise. In
one case (``pion-like'', figure~\ref{fig:synthetic_pion}) the
amplitude of the noise scaled linearly with the signal, in the other
(``rho-like'', figure~\ref{fig:synthetic_rho}) the noise amplitude was
kept constant. 
It can be seen that the ground state mass and the mass of the first
excited state are extracted with high reliability, while for the
higher states good estimates are obtained.
The increased error in the excited states is largely due to the noise
component in the created data which makes it impossible to adequately
resolve those states and not to a shortcoming in the algorithm itself.

\begin{figure}
\includegraphics[height=\textwidth,angle=270,keepaspectratio=]{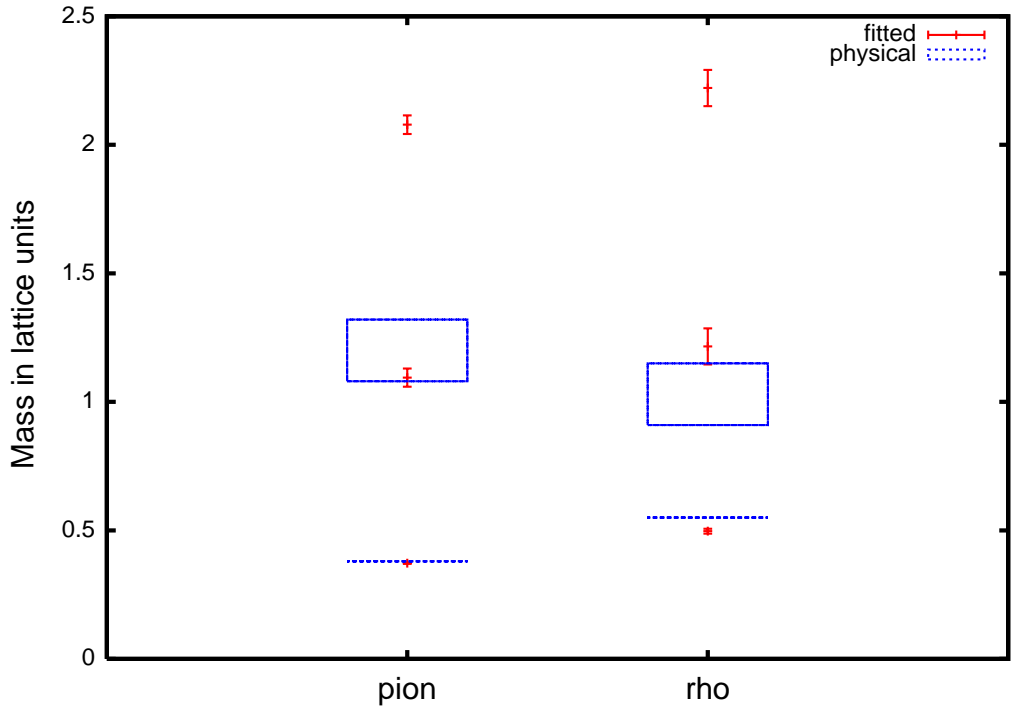}
\caption{Result of a fit to an actual meson correlator from a quenched
  simulation using Wilson quarks. The simulation ($N_{conf}=100$) was
  run on a $16^3\times 32$ lattice at $\beta=6.0$, $\kappa=0.154$.
  The red points and error bars are the results of the fit; the blue
  bands are the approximate masses of the continuum states
  corresponding to the lowest and first excited states in the fitted
  channels extrapolated to the quark masses employed in the simulation.
  The largest fitted state does not correspond to a known continuum state
  and may be a lattice artifact.\vspace{3ex}}
\label{fig:real_data}
\end{figure}

The running time for our \python\ implementation of each of the two
bootstrapped fits was on the order of one to two hours on a
single-processor Pentium 4 workstation; implementing the same
algorithm in a compiled language could cut this runtime down
further. It should be stressed that the evolutionary fit works
robustly without any human intervention, thus saving valuable human
time at the expense of a moderate increase in computer time when
compared to more conventional excited state fitting methods that often
require a larger degree of user intervention.

Finally, in figure~\ref{fig:real_data} we show the results from fits to
actual pion and rho correlators from a quenched simulation using
Wilson quarks, demonstrating the ability of our program to deal
with real lattice data.

\section{Evolutionary Fitting of Multiple Correlators}
\label{sec_evmulti}

In this section we discuss the generalizations of the evolutionary
fitting algorithm required for fitting multiple datasets simultaneously. 
A more sophisticated genotype is required in this case, since
now the same energy $E_n$ can occur throughout some subset of the datasets.
Hence, the fit for dataset number $i$ will now be represented by a list of
$n^{(i)}_{max}$ coefficients $(Z_n^{(i)},I_n^{(i)})$, $n =
1,\ldots,n^{(i)}_{max}$ where $I\in\{1,\ldots,m_{max}\}$ is an integer
index into a list of $m_{max}$ energy states $E_m$ indicating to which
state the coefficient is associated.\footnote{The stored integer is
interpreted modulo $m_{max}$ to ensure it maps to an actual energy state.}
The list $E_1 \ldots E_{m_{max}}$ is common to all datasets.

In summary, for a fit of $i_{max}$ datasets the complete genotype is of
the form:
\begin{eqnarray}
  \label{eq:multidatasetfit}
  \lefteqn{\mathrm{Fit\ Genotype}} \nonumber \\
  & = & ( \mathrm{Dataset\ coefficients},\mathrm{Mass\ list} ) \nonumber \\
  & = & ( (\mathrm{Dataset\ 1\ coefficients},\ldots,\mathrm{Dataset\ }i_{max}\mathrm{\ coefficients}),\mathrm{Mass\ list}  ) 
\end{eqnarray}
with
\begin{eqnarray}
 \mathrm{Dataset\ }i\mathrm{\ coefficients} & = & ( (Z_1^{(i)},I_1^{(i)}), \ldots, (Z_{n_{max}^{(i)}}^{(i)},I_{n_{max}^{(i)}}^{(i)})) \nonumber \\
  \mathrm{Mass\ list} & = & (E_1, \ldots ,E_{m_{max}})\ .
\end{eqnarray}
Assuming the datasets are not correlated in any way, the new $\chi^2$
is simply the sum of terms of the form in
equation~(\ref{eq:chi2_single}), one per dataset.  The number
of degrees of freedom changes from the form of equation~(\ref{eq:ndof_single}) to: 
\begin{equation}
\label{eq:chi2_multi}
n_{dof}(G)=n_{data} - m_{max}-\sum_{i=1}^{i_{max}}n_{max}^{(i)}
\end{equation}
where $n_{data}$ now counts all included timesteps in the fit of all
datasets.

The presence of integer variables in the multi-dataset genotype
requires some adaptations to the breeding algorithm used. For
practical reasons, it is advantageous to use a fixed-length integer
implementation with $n$-bit integers, where $2^n$ is the maximum
number of distinct states we will allow in our fit. Crossover can be
performed by exchanging the first $m$ ($0\le m\le n$) bits of two
integers. Mutations are implemented by flipping each bit of the
integer with some fixed probability.

The genotype~(\ref{eq:multidatasetfit}) has a hierarchical structure
as a list of lists ultimately containing floating point or integer
numbers. Mutation and crossover can therefore be structured by
recursing through these list structures, applying appropriate mutation
and crossover operations at each level. The lists can be distinguished
by the number (fixed or variable) and type (homogeneous or
heterogeneous) of their elements, and by whether they are ordered or
unordered. Elementwise mutation can
be done on any kind of list. Lists of variable length can also be
mutated by removing elements or adding a random new element, and
ordered lists may be mutated by permuting their elements. Likewise for
crossover, building two new lists by picking from the elements of the
parent lists in order can be done for any list. Homogeneous
lists allow more general subsets of the parents' elements to be chosen.
Elementwise crossover may also be done, in order for heterogeneous lists or
with random pairs of the parents' elements in the homogeneous case.
All of the different mutation and crossover operations that are
possible in each case can have different probabilities assigned to
them and for lists of variable length, a range of valid list lengths
may be specified.

In addition to these generic genetic operations, the multi-dataset
fitting problem benefits from some operations specific to its
structure.  One notes that the genotype of
equation~(\ref{eq:multidatasetfit}) allows
for multiple representations of the same fitting function, 
because the coefficient integers are taken modulo the number of masses
in the fit, because either the masses or the indices to them
may be in different orders, and because some masses may be unused or
referred to multiple times for a single dataset. This degeneracy can
have an adverse effect on the convergence of the algorithm. To
encourage the algorithm to work toward a single representation of the
solution, we employ a reduction mutation which sorts the mass
list and removes unused masses.  It also combines coefficients
pointing to the same mass, places coefficient indices uniquely within
the range $1,\ldots,m_{max}$, and orders them within each dataset by
the associated mass index. In order to favor this operation, as well
as any other mutation that may reduce redundancies, it is beneficial
to count all masses and coefficients, whether redundant or not, in the
count of the degrees of freedom in equation~(\ref{eq:chi2_multi}).

As in the single-correlator case, interspersing local optimization
steps using the Levenberg-Marquardt method with the other mutations
was found to be useful in accelerating convergence. Limiting the
mutation to a fixed number of steps of the Levenberg-Marquardt
algorithm and making it a relatively improbable mutation keeps the
overall time evolution reasonable.

Special care needs to be taken in the multi-dataset case when adding
or removing masses, since simply dropping a mass from the list will
mean that any coefficients belonging to the dropped mass will now
become associated with a different mass, which tends to have a
catastrophic effect on fitness, especially for more evolved
genotypes. A careful generalization of the way in which amplitudes
are redistributed when changing the number of masses in the
single-correlator case is therefore necessary. This is facilitated by
first applying the reduction operation mentioned above before adding
or removing a mass from the mass list. When removing a mass, one then
loops through the dataset functions to see which ones have a
coefficient corresponding to the removed mass. For any that do, the
coefficient is deleted and its amplitude is redistributed between
neighboring masses, either by increasing the coefficients for the
closest masses already having coefficients in that dataset function,
or by adding a new coefficient for an unrepresented neighboring
mass. When adding a mass, at least one of the datasets receives a new
coefficient associated with it, and the coefficients of the other
masses in those datasets are lowered accordingly. As a final step, a
Levenberg-Marquardt optimization step is carried out in order to give
the new mutant a better chance of survival. There is obviously a
tradeoff between the probability of such relatively expensive
mutations and the maximum size of the population that can be
employed.

\begin{figure}
  \includegraphics[width=\textwidth,angle=0,keepaspectratio=]{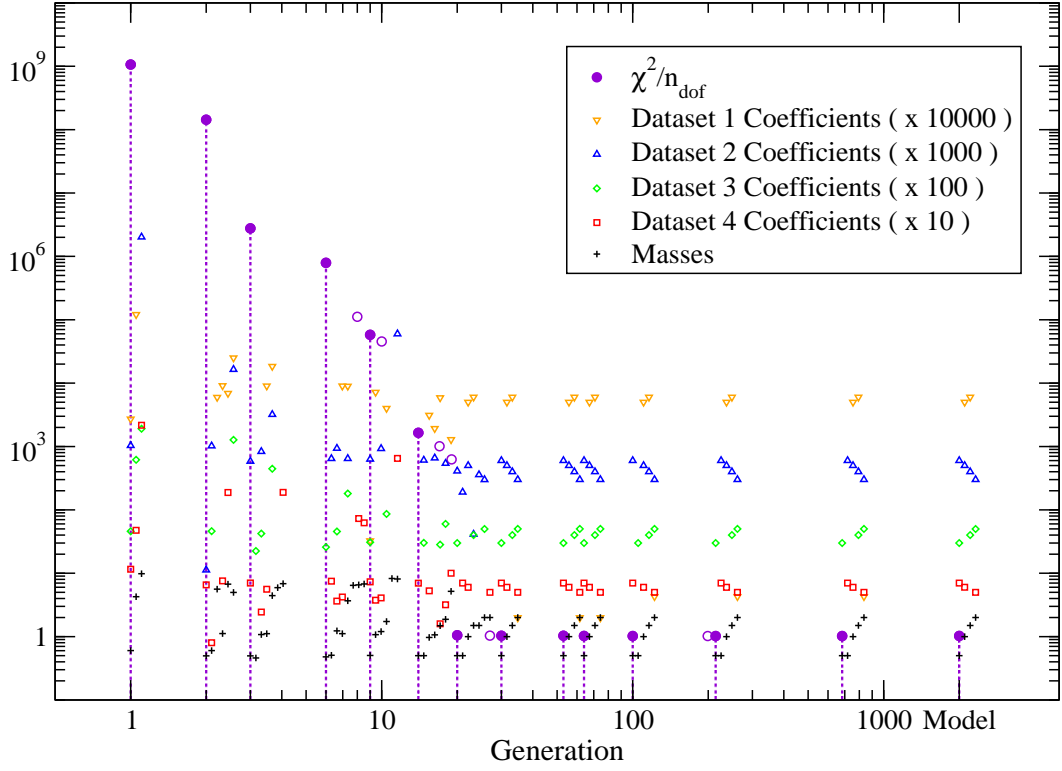}
  \caption{Best fit genotype and its $\chi^2/n_{dof}$ of each
    generation for a multi-dataset fit are shown.  Four datasets were
    constructed to contain four masses with each dataset having
    coefficients for two to four of the masses.  The model fit that
    generated the data is shown on the right.  Each fit is displaced
    horizontally with different masses in separate columns with their
    corresponding coefficients in each of the datasets vertically
    aligned above them.  The latter have been multiplied by powers of
    $10$ for display purposes.  The left side of the plot shows the
    results of the evolutionary algorithm, starting with the best fit
    of the first generation.  Only generations with an improvement in
    the best fit genotype are plotted.  For clarity a few of these
    fits have only their $\chi^2/n_{dof}$ displayed (open circles) to
    ensure that all data between subsequent drop lines corresponds to
    a single fit.\vspace{3ex}}
  \label{fig:synthetic_multidataset}
\end{figure}

As a test of the efficiency of the evolutionary algorithm in the
multi-dataset case we have run fits of synthetic data. The result of
such a test is shown in figure~\ref{fig:synthetic_multidataset}.
Four hypothetical diagonal meson correlators were constructed to
contain four masses among them. 
One correlator had coefficients for only two of the masses, two had
coefficients for three of the masses, and one had coefficients for all
four masses.  The four datasets, with $48$ timesteps each,
were modified by adding Gaussian noise, the amplitude of which was
chosen small enough to allow statistical discrimination of all states. 
A Levenberg-Marquardt fit to the synthetic data, using the masses and
coefficients used to generate the data as the starting point,
is shown in the last column of
figure~\ref{fig:synthetic_multidataset}.
As expected, the parameters shifted only marginally
from their ideal values and $\chi^2/n_{dof}$ for the fit is very
close to $1$.

The multi-dataset algorithm was then run on the synthetic data.  It
was restricted to fit genotypes containing a maximum of eight masses
with positive masses of value less than $10$.  Each dataset function
was allowed to have up to six coefficients whose values were only
restricted to being positive.  Four islands were used, each containing
$120$ individuals (specifically
$N_{elite}=N_{diversity}=5$,~$N_{mutants}=20$).
Figure~\ref{fig:synthetic_multidataset} displays 
the best fit genotype from each generation along with its
$\chi^2/n_{dof}$. It is evident that the algorithm quickly succeeds at
finding a good fit to the data.  By generation~$20$ one has a fit with
good $\chi^2/n_{dof}$ and characteristics almost identical to the
model function, having the same number of masses and corresponding
coefficients, except for the addition of a small (four orders of
magnitude lower) coefficient added in the first dataset for the
highest mass.  Generation $64$ shows that this functional form with an
additional coefficient actually improves the fit, having a lower
$\chi^2/n_{dof}$ than the optimized model function.  Generations $100$
onward \emph{improve} even further in the fit by bifurcating the
lowest mass to produce a fit of ever so slightly \emph{greater} probability
than the actual model function used to generate the data. This was the
final result as of generation~$2000$. The total run was performed
overnight on a single-processor Pentium 4 workstation.

Overall one notices that the algorithm adds masses initially as
required before proceeding to coalesce masses and combine coefficients
to improve the fit.  The reduction mutation serves to encourage
an ordering of masses from lowest to highest but generations $2$ to
$10$ show it is not strictly required.

\begin{figure}
\includegraphics[width=\textwidth,angle=0,keepaspectratio=]{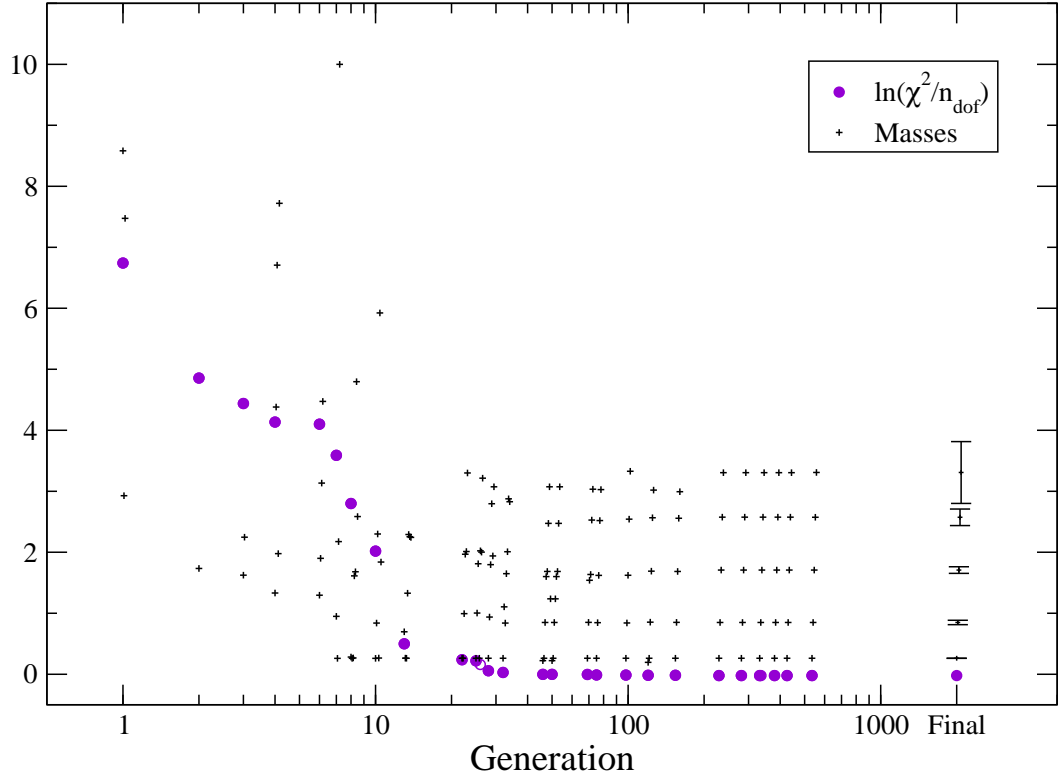}
\caption{A simultaneous fit to actual data of eight diagonal pion (i.e.
$\Lambda^{PC}=A_1^{-+}$) correlators from a quenched simulation using
Wilson quarks ($\beta=6.0$, $\kappa=0.1554$, $20^3\times 48$, $600$
configurations) is shown; the eight operators used are from
\cite{Harnett:2006fp}. Only the energies 
of the fittest organism of each generation are shown; the coefficients
in each dataset, a further $28$ parameters in the final fit, are not.
The last column depicts the best fit found with bootstrap errors
produced via Levenberg-Marquardt fits with its fixed functional form.
Also shown (circles) is (the logarithm of) $\chi^2/n_{dof}$ of the best
genotype, which indicates that by generation $25$ one has
technically a good fit ($\chi^2/n_{dof}\approx 1$).\vspace{3ex}}
\label{fig:multipion}
\end{figure}

We are currently employing this program in a forthcoming analysis
\cite{Petry:forthcoming}
of the meson spectrum of twisted-mass QCD
\cite{Frezzotti:2000nk}
based on the representation theory of the octahedral group with
generalized parity
\cite{Harnett:2006fp}.
In figure~\ref{fig:multipion}, we show the results of a fit to actual
data from the Wilson QCD action in the pion channel.
The algorithmic parameters were the same as
for the previous fit. Again, one observes that initially the
number of states found fluctuates considerably, with states being
added to improve the fit.  At some point, too many states which are 
nearly degenerate have been introduced, and the population culls
unnecessary bifurcations. One observes that by generation $200$ the
energy states have been largely found, further improvements occurring
within the coefficients. The fitness of the best genotype traces this
behavior.\footnote{As an aside we note that for this
fit and the fit to synthetic data shown in
figure~\ref{fig:synthetic_multidataset},
the restriction $E_n<10$ was initially imposed with an eye to restricting
ourselves to physical masses and thereby preventing the occurrence of
``runaway'' solutions commonly encountered in this fitting task.  We
subsequently found this restriction to be unnecessary. We conjecture
that the problem with runaway solutions may be due to trying to fit a
poor functional form to the data, something which our algorithm
avoids, thus removing the need for such a constraint.  Subsequently,
we merely constrain parameters to be positive.}

\begin{figure}
\includegraphics[width=\textwidth,angle=0,keepaspectratio=]{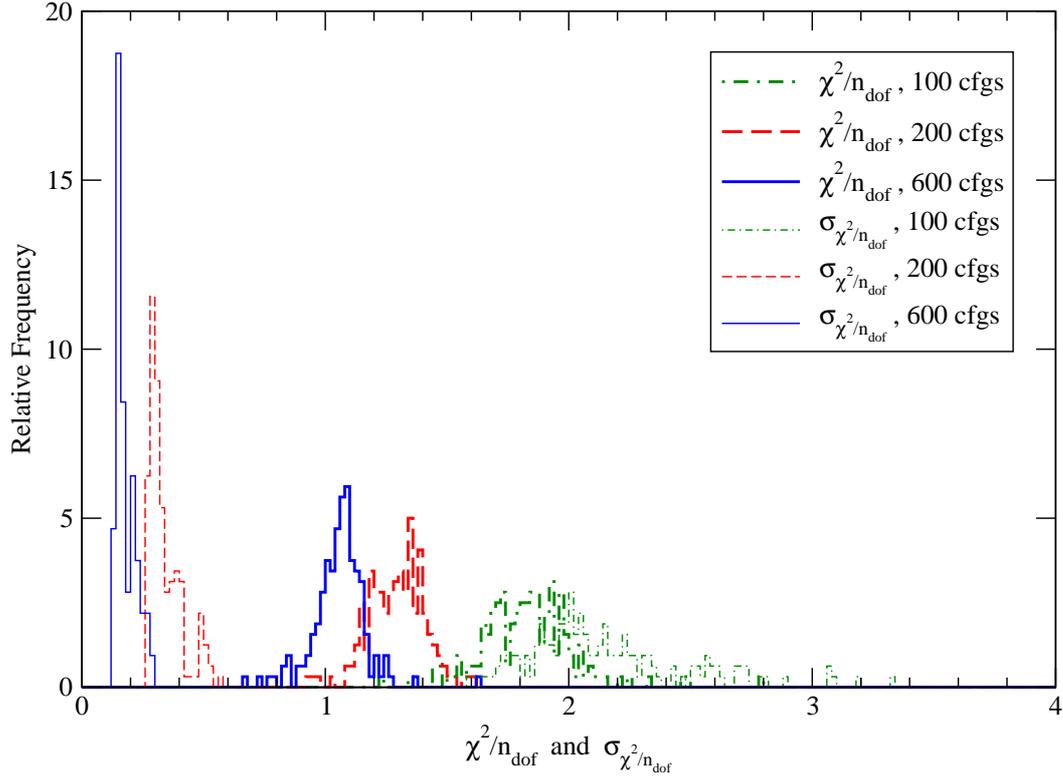}
\caption{Shown are normalized histograms of $\chi^2/n_{dof}$ (thick
lines) of fits to $160$ meson quantum channels containing between $2$ 
and $16$ correlators each. 
For each channel the evolutionary algorithm was run three times:
using all ($N_{config}=600$ configurations), one third
($N_{config}=200$), and one sixth ($N_{config}=100$) of the lattice
data.  For the full data set, the fitness distribution peaks sharply
around $1.0$, but as the amount of data included drops,the certainty
with which states are resolved falls and the optimal fit becomes
poorer.  Bootstrap errors of the fitness, $\sigma_{\chi^2/n_{dof}}$,
are also shown (thin lines). The stability of the fit is seen to
decrease in the same way as the overall $\chi^2/n_{dof}$ as the
quality of the data declines.\vspace{3ex}}
\label{fig:cfgcompare}
\end{figure}

To get an impression of how well the algorithm converges, we show
histograms of $\chi^2/n_{dof}$ values of the final best fit for
160 twisted-mass meson channels in figure~\ref{fig:cfgcompare}.
The 160 channels
correspond to four quark masses (the mesons consist of mass-degenerate
quarks and antiquarks), two particle types (charged or neutral), and
$20$ octahedral irreducible representations ($\Lambda^{PC}$).
The $400$ operators used ($16$ local and $384$ extended) are detailed
in
\cite{Harnett:2006fp}.
Diagonal correlators were obtained for all of these operators,
encompassing all channels. For operators in representations of
dimension greater than one, all the correlators  were row-averaged
before fitting. The quenched simulation used degenerate twisted mass
quarks with quark and link smearing of operators on a $20^3\times 48$
lattice at $\beta=6.0, m \sim m_s, m_s/2, m_s/3, m_s/6$ (where $m_s$
is the physical strange quark mass); see
\cite{AbdelRehim:2006qu,AbdelRehim:2005gz,AbdelRehim:2006ve}
for further details.

For each channel the evolutionary algorithm was run three times: first
using all of the data ($N_{config}=600$ configurations), then one
third of it ($N_{config}=200$), and finally one sixth of it
($N_{config}=100$).  The fitness distribution for the full data set is
excellent with $\chi^2/n_{dof}$ distributed locally about $1.0$, clearly
demonstrating the algorithm is robust in its ability to find a good
fit, when the latter exists.   As the amount of data included drops,
however, the certainty with which states are resolved falls and the
optimal fit becomes poorer. Fits required a minimum of $600$
generations, and ran for up to $1200$ generations if improvement still
occurred in the best organism.  The shown fits are for correlators
with smeared operators. Unsmeared correlators (not shown) of the same
operators which exhibit a greater number of excited states were also
fit for the $600$ configuration case with no appreciable difference
from the histogram of the smeared one.

\begin{figure}
\includegraphics[width=\textwidth,angle=0,keepaspectratio]{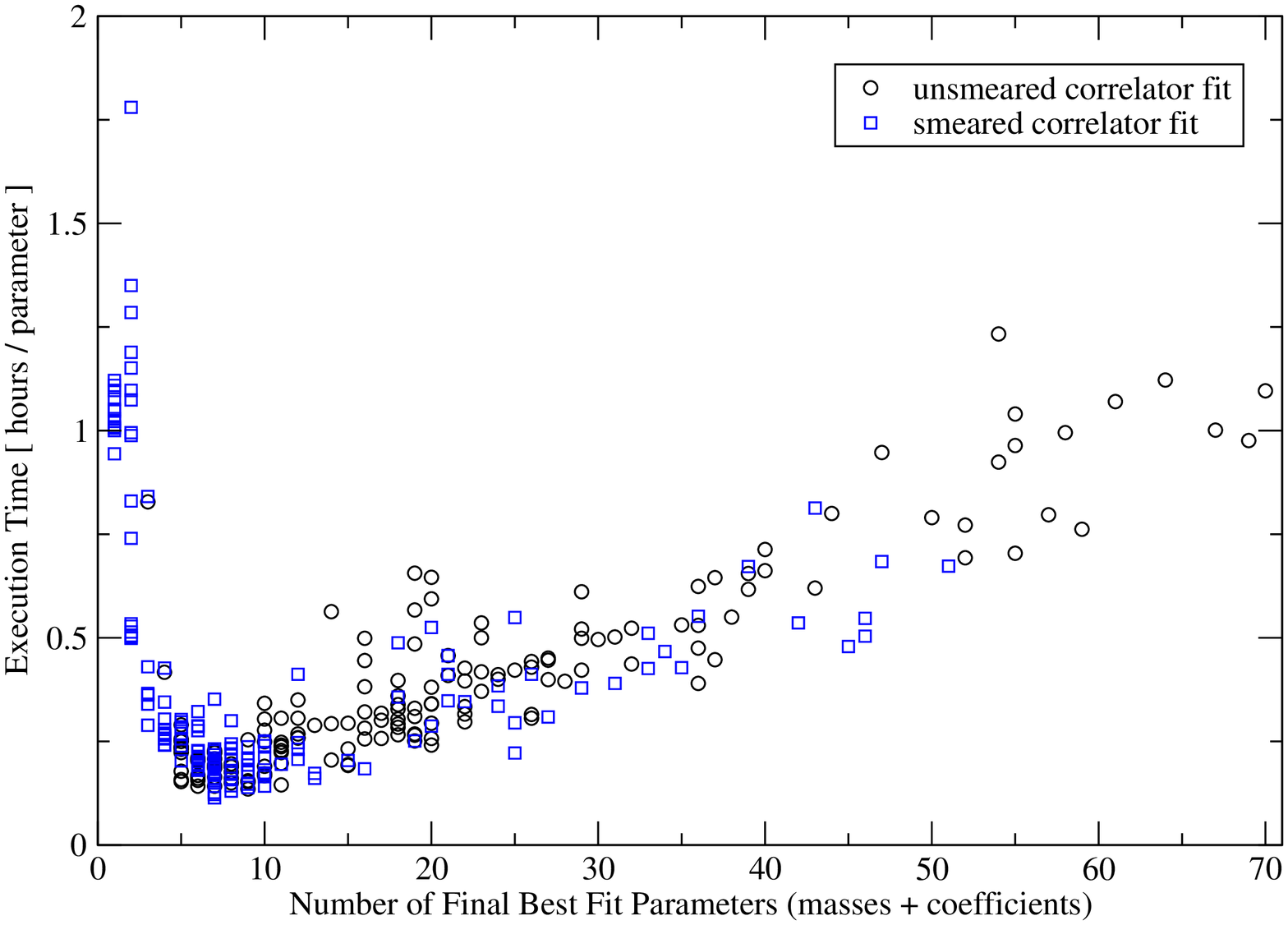}
\caption{Shown is the execution time (on a single Intel Xeon 3.06 GHz
  CPU) per parameter as a function of the number of parameters in the
  final best fit.  The latter includes each of the masses as well as
  the coefficients on each dataset for the given fit.  The number of
  generations for all fits shown was fixed at 600, by which time all
  fits were already well converged.  Smeared correlators (squares)
  exhibit a smaller number of fitted parameters than unsmeared ones
  (circles), which is expected since smearing suppresses contributions
  from excited states.\vspace{3ex}}
\label{fig:timeperparameter}
\end{figure}

As a measure of stability of these fits, bootstrap errors of the fitness,
$\sigma_{\chi^2/n_{dof}}$, have been calculated for each fit and their
histograms are also shown. Specifically, for each
fit, $3\times N_{config}$ bootstrap configurations of the data were made
and the functional forms of the best fits were fit to them using a
Levenberg-Marquardt routine whose initial values were those of the best
fit. For the full $600$ configurations one observes the variation in
the goodness of the fit is small ($\sigma_{\chi^2/n_{dof}}<0.3$). Since
good fits occur in essentially all of the bootstrap configurations, this
increases confidence that the spectrum found is accurate.  The same is
largely true for $200$ configurations as well
($\sigma_{\chi^2/n_{dof}}<0.6$), but by the time the data set is reduced
to only $100$ configurations the error in $\chi^2/n_{dof}$ is as wide as
the value itself, indicating along with the fit histogram, that the
amount of data is insufficient for solution of the problem.

Finally, in figure~\ref{fig:timeperparameter} we show the runtime per
parameter as a function of the total number of parameters (masses and
coefficients in all datasets) in the final fit. Overall, we find that
the dominant contribution to runtime comes from the number of
coefficients, which in turn is largely proportional to the number of
correlators used for the fit, as might be expected from a naive
runtime analysis.

\section{Alternatives and Variations}
\label{sec_altvar}

The specific implementation described in the previous section is just
one of many possible ways to approach the problem of extracting
excited state energies using evolutionary algorithms. In this section we
wish to point out some alternatives to our current implementation,
many of which we are actively exploring at the moment.

The previously described algorithm used a genetic encoding based on
real numbers. This offers the advantage of being a straightforward
representation, as well as the ability to interject local optimization
steps in a natural manner. An alternative would be to use an
integer-based representation of real parameters instead;
the advantage in this case would
be the ability to use bit-based mutation and crossover operations
instead of our Gaussian mutations and interpolating crossover. The
bit-based operations, besides likely being faster in most cases, are
better understood in terms of rigorous theorems regarding global
convergence properties (such as from schema theory
\cite{Goldberg:1989,Whitley:1994}),
but do not offer the possibility for easy mixing with local
optimization.

Using straight elitism as the selection method has been found to be
favorable in the case of problems with a real-number genetic code
\cite{Allanach:2004my},
but introduces a certain risk of premature convergence if the elite
should happen to cluster around a local optimum. This risk is
significantly reduced by the addition of random survivors and our use
of an island-based ecosystem, but other, more sophisticated, selection
methods could help to further eliminate any remaining bias. Another
disadvantage of elitist selection is that it requires a full sort of
the organisms by fitness in each generation, which makes up a fair
part of our implementation's computational cost. Other selection
methods manage to avoid this requirement, which could lead to further
gains in speed. The specific mutation schedule implemented in our
algorithm is also to be seen as just one example out of many that are
possible. In some cases, it may be more favorable to mutate each
parameter separately instead of mutating all parameters at once.

Our evolutionary algorithm depends on a number of parameters, such as
the rate for different kinds of mutations, their dependence on
$\chi^2/n_{dof}$, and the size of the breeding pool. In the current
implementation, these parameters have been set to reasonable values by
hand. For a more highly optimized implementation, these parameters
should be tuned to values that tend to give the fastest rate of
convergence towards the true optimum; in principle, such tuning itself
could be done by means of another evolutionary algorithm, although
that approach might prove to be fairly expensive computationally.

Evolutionary multi-modal algorithms (see
\cite{1144200}
and refs. therein) are able to find not only absolute
extrema but relative extrema as well.  ``Niching'' and
``diversity preservation'' algorithms have been devised that dissuade
too many elements of the population from going after the single best
solution, thereby finding not only the best solution but other good
solutions. This could be useful in the context of fitting since these
algorithms can produce the best fit as well
as other fits that lie in relative extrema that are perhaps comparable
with the best fit.  Comparison of such fits would give the researcher
a better feel for the uniqueness and likelihood of the functional form
of the solution found.

Several improvements might be made for the simultaneous fitting of
multiple correlators.  Obviously it is of value to cull from the fit
any datasets which clearly only contain noise.  As well, if fitting a
large number of datasets it could be of value to partition the
datasets and find viable fits on each subset first.  One could then
merge these populations into genotypes suitable for the entire dataset
by stitching together fits from each subset.  This could be done
crudely by just putting the masses back to back, or one could
implement some algorithm which tried to find common masses at this
point between two genotypes being merged in some systematic fashion.
Running the algorithm on these new datasets would then optimize these
fits globally, presumably by coalescing common masses across the
subsets that are statistically equal to improve the final fit.

The local optimization steps used as mutations in our algorithm could
be rendered more efficient by employing techniques that exploit the
partial linearity of the functional form of equation
(\ref{eqn:thform}) by separating the linear and non-linear variables
\cite{Golub:1973}.

There is also the possibility to combine the variational method with
an evolutionary fitting algorithm. To do this, one could diagonalize
the correlator matrix as usual with the variational method, and then
use the evolutionary algorithm to fit the resulting diagonal
correlators. In this way, any remaining mixing between the optimized
operators could be detected and quantified, while at the same time
reducing the number of correlators to be fit.

An alternative to evolutionary algorithms, which we have not
investigated so far, might be the use of Markov Chain Monte Carlo
optimization methods such as simulated annealing
\cite{Kirkpatrick:1983},
which share evolutionary algorithms' ability to accommodate discrete
changes in functional form.

\section{Conclusions}
\label{sec_conc}

Evolutionary fitting methods provide an interesting and useful
addition to the lattice field theorist's data analysis
toolkit. Especially when combined with other well-known and
well-tested fitting methods, evolutionary fitting can help to extract
information from simulation data without having to impose any external
constraints, such as Bayesian priors. Evolutionary methods allow one
instead to extract all of this information from the data themselves by
harnessing the globally optimizing nature of evolving systems. This is
particularly true in the case of discrete parameters such as the
number of states to fit, which are hard to determine using more
conventional methods.

We have demonstrated a working method for the extraction of excited
state masses from lattice QCD correlators using evolutionary
fitting. We believe that evolutionary fitting algorithms have significant
potential as a data analysis method in lattice QCD, and that further
investigation in this direction is warranted.

\begin{ack}
We thank the reviewer for pointing out the potential of Markov Chain
Monte Carlo Methods as a possible alternative to evolutionary
algorithms, and George Fleming for bringing reference
\cite{Golub:1973} to our attention.

We also thank Richard M. Woloshyn for providing gauge field
configurations and propagators
\cite{AbdelRehim:2005gz,AbdelRehim:2006ve}.

This work was supported in part by the Natural Sciences and
Engineering Research Council (NSERC) of Canada, the Canada Research
Chairs Program, and the Government of Saskatchewan.
\end{ack}

%
% Bibliography
%

\end{document}